\documentclass{aa}
\usepackage{graphicx}
\usepackage{epsfig}
\usepackage{float}
\usepackage{amsmath}
\usepackage{color}
\usepackage{amssymb}
\usepackage{amsfonts}
\usepackage{units}
\usepackage{bm}
\usepackage{txfonts}
\begin{document}

        \title{Interpreting time-integrated polarization data of gamma-ray burst prompt emission}
        \author{R.Y. Guan
                \inst{1}
                \and
                M.X. Lan\inst{1}
        }
        \institute{Center for Theoretical Physics and College of Physics, Jilin University,
                 Changchun 130012, China\\
                \email{lanmixiang@jlu.edu.cn}
        }
        \date{}
\abstract
{}
{With the accumulation of polarization data in the gamma-ray burst (GRB) prompt phase, polarization models can be tested.} 
{We predicted the time-integrated polarizations of 37 GRBs with polarization observation. We used their observed spectral parameters to do this. In the model, the emission mechanism is synchrotron radiation, and the magnetic field configuration in the emission region was assumed to be large-scale ordered. Therefore, the predicted polarization degrees (PDs) are upper limits.}   
{For most GRBs detected by the Gamma-ray Burst Polarimeter (GAP), POLAR, and AstroSat, the predicted PD can match the corresponding observed PD. Hence the synchrotron-emission model in a large-scale ordered magnetic field can interpret both the moderately low PDs ($\sim10\%$) detected by POLAR and relatively high PDs ($\sim45\%$) observed by GAP and AstroSat well. Therefore, the magnetic fields in these GRB prompt phases or at least during the peak times are dominated by the ordered component. However, the predicted PDs of GRB 110721A observed by GAP and GRB 180427A observed by AstroSat are both lower than the observed values. Because the synchrotron emission in an ordered magnetic field predicts the upper-limit of the PD for the synchrotron-emission models, PD observations of the two bursts challenge the synchrotron-emission model. Then we predict the PDs of the High-energy Polarimetry Detector (HPD) and Low-energy Polarimetry Detector (LPD) on board the upcoming POLAR-2. In the synchrotron-emission models, the concentrated PD values of the GRBs detected by HPD will be higher than the LPD, which might be different from the predictions of the dissipative photosphere model. Therefore, more accurate multiband polarization observations are highly desired to test models of the GRB prompt phase.}   
{}

\keywords{polarization -- gamma-ray burst: general -- radiation mechanisms: non-thermal -- methods: numerical -- magnetic fields
}
\maketitle
%

\section{Introduction}

Gamma-ray bursts (GRBs) are the most violent high-energy explosions in the Universe. GRBs were divided into two categories, long and short GRBs, based on a rough duration separation of about 2 seconds. GRB spectra are nonthermal, which can be described by a broken power law with a smooth joint, known as the Band function \citep{Band1993}. The spectrum integrated over the GRB duration can empirically be described by a function with a peak in the $\nu f_{\nu}$ spectrum, and the peak energy is defined as $E_{p,obs}$. For the low-energy spectral index $\alpha$, the typical value for long GRBs is $\alpha\sim-0.92$, and short GRBs have a harder low-energy spectral index $\alpha\sim-0.50$ \citep{Nava2011}.

Gamma-ray polarization measurements of the prompt emission of GRBs have profound implications for our understanding of the unknown magnetic field configuration and emission mechanism of GRBs prompt emission. With the development of polarimetry, increasingly more GRBs have been measured and can be used for a statistical analysis. Therefore, constraints on the underlying models can be provided \citep{Toma2009}. For the GRB prompt phase, there are two possible emission mechanism, synchrotron radiation and inverse Compton scattering \citep{Chand2018, Lazzati2004}. Although several thousands of GRBs have been observed to date, few of these have reported polarization detections. The polarization degrees (PDs) of GRB prompt emissions in the measurements so far vary strongly.

The Gamma-ray Burst Polarimeter (GAP) has observed PD values of GRBs 100826A, 110301A, and 110721A, which suggests that GRB prompt emissions are highly polarized \citep{Yonetoku2011, Yonetoku2012}. Then an increasing number of polarimeters became operational. \citet{Chattopadhyay2022} published their updated polarization observational results of 20 GRBs recently, which are the brightest GRBs detected by the Cadmium Zinc Telluride Imager (CZTI) on board AstroSat. The renewed AstroSat data show that most of the bright GRBs are relatively highly polarized (with a typical PD value of $\sim45\%$) in the energy range of 100 keV$-$600 keV, different from that of their former results with the high polarizations (typical PD was around $60\%$) in energy range of 100 keV$-$350 keV \citep{Chattopadhyay2019, Chand2019, Gupta2022}.

POLAR is a polarimeter with a comparable energy range as CZTI, which was launched as part of the China Tiangong-2 space laboratory in September 2016. The detection energy range of POLAR is 50 keV$-$500 keV. During its approximately six months of operation, a total of 55 GRBs were detected \citep{Xiong2017}. Polarization measurements of 5 of these 55 GRBs were reported first, and the results show that they are less polarized than predicted by some popular models \citep{Zhang2019}. Moderate levels of linear polarization were also found in subsequent reports, and the polarization measurements of 9 other GRBs were published next \citep{Kole2020}. Despite the great efforts that have been made in gamma-ray polarimetry, there are still large errors in the current data, which allow us to present only preliminary constraints on the various models of the GRB prompt phase. It is encouraging that more detailed polarization measurements will become available from forthcoming missions such as POLAR-2 \citep{POLAR2}, which will help us to understand the magnetic field configuration and emission mechanism of GRBs.

In this paper, we have numerically calculated \citep{Toma2009} the ranges of theoretical PDs of 37 GRBs detected by GAP, AstroSat, and POLAR based on the values of the observed spectral parameters. The paper is arranged as follows. In Section 2 we present our data. The model and numerical results are described in Section 3. Finally, we give our conclusions and discussion in Section 4.

\section{Data list}

\begin{table*}[!htbp]

        \caption{Spectral parameters and polarization properties of the three GRBs observed with GAP}
        \label{tab1: GAP}
        \begin{center}
                \centering
                \begin{tabular}{cccccccc}
            \hline\hline\noalign{\smallskip}
                        GRB  &  $PD_{obs}$(\%)&$\alpha_{s}$&$\beta_{s}$&$E_{p,obs}$(keV)&Instrument(Spectrum)&$z$&\\
                        \hline\noalign{\smallskip}
                        100826A&$27_{-11}^{+11}$&$-0.19_{-0.01}^{+0.01}$&$0.92_{-0.02}^{+0.02}$&$263.25_{-7.84}^{+7.84}$&Fermi-GBM&-&\\[5pt]
                        110301A&$70_{-22}^{+22}$&$-0.10_{-0.02}^{+0.02}$&$1.67_{-0.05}^{+0.05}$&$102.28_{-1.82}^{+1.82}$&Fermi-GBM&-&\\[5pt]
                        110721A&$84_{-28}^{+16}$&$0.03_{-0.02}^{+0.02}$&$0.78_{-0.03}^{+0.03}$&$465.19_{-38.66}^{+38.66}$&Fermi-GBM&$0.382$&\\\noalign{\smallskip}
                        \hline
                \end{tabular}
        \end{center}
\end{table*}

\begin{table*}[htbp]
        \caption{Spectral parameters and polarization properties of the 14 GRBs observed with POLAR}
        \label{tab2: POLAR}
        \begin{center}
                \begin{tabular}{c c c c c c c c c c}
                        \hline\hline\noalign{\smallskip}
                        GRB  &  $PD_{obs}$(\%)&$\alpha_{s}$&$\beta_{s}$&$E_{p,obs}$(keV)&Instrument(Spectrum)&$z$&\\
                        \hline\noalign{\smallskip}
                        161203A&$16_{-15}^{+29}$&$-1.13_{-0.27}^{+0.25}$&$2.41_{-0.39}^{+0.46}$&$344_{-12}^{+19}$&$^{*}$&-&\\[5pt]
                        161217C&$21_{-16}^{+30}$&$0.08_{-0.43}^{+0.25}$&$1.76_{-0.36}^{+0.61}$&$143_{-34}^{+32}$&$^{*}$&-&\\[5pt]
                        161218A&$7.0_{-7.0}^{+10.7}$&$-0.72_{-0.25}^{+0.21}$&$2.40_{-0.43}^{+1.17}$&$128_{-8}^{+8}$&Konus-Wind&-&\\[5pt]
                        161218B&$13_{-13}^{+28}$&$-0.52_{-0.01}^{+0.01}$&$1.93_{-0.10}^{+0.10}$&$209.67_{-3.00}^{+3.00}$&Fermi-GBM&-&\\[5pt]
                        161229A&$17_{-13}^{+24}$&$-0.36_{-0.03}^{+0.03}$&$2.07_{-0.72}^{+1.49}$&$339_{-14}^{+12}$&$^{*}$&-&\\[5pt]
                        170101A&$6.3_{-6.3}^{+10.8}$&$0.44_{-0.17}^{+0.13}$&$1.49_{-0.23}^{+0.65}$&$123_{-21}^{+23}$&Konus-Wind&-&\\[5pt]
                        170101B&$60_{-36}^{+24}$&$-0.43_{-0.06}^{+0.06}$&$1.23_{-0.12}^{+0.12}$&$206.52_{-12.75}^{+12.75}$&Fermi-GBM&-&\\[5pt]
                        170114A&$10.1_{-7.4}^{+10.5}$&$-0.17_{-0.05}^{+0.05}$&$1.04_{-0.09}^{+0.09}$&$230.15_{-21.03}^{+21.03}$&Fermi-GBM&-&\\[5pt]
                        170127C&$9.9_{-8.4}^{+19.3}$&$0.14_{-0.22}^{+0.21}$&$2.1_{-0.6}^{+0.6}$&$1500_{-900}^{+800}$&$^{*}$&-&\\[5pt]
                        170206A&$13.5_{-8.6}^{+7.4}$&$-0.72_{-0.04}^{+0.04}$&$1.55_{-0.12}^{+0.12}$&$341_{-13}^{+13}$&Fermi-GBM&-&\\[5pt]
                        170207A&$5.9_{-5.9}^{+9.6}$&$-0.14_{-0.06}^{+0.06}$&$1.63_{-0.26}^{+0.84}$&$394_{-33}^{+42}$&Konus-Wind&-&\\[5pt]
                        170210A&$11.4_{-9.7}^{+35.7}$&$-0.10_{-0.02}^{+0.02}$&$1.28_{-0.08}^{+0.08}$&$361.5_{-14.1}^{+14.1}$&Fermi-GBM&-&\\[5pt]
                        170305A&$40_{-25}^{+25}$&$-0.58_{-0.13}^{+0.13}$&$1.06_{-0.13}^{+0.13}$&$233_{-35}^{+35}$&Fermi-GBM&-&\\[5pt]
                        170320A&$18_{-18}^{+32}$&$-0.76_{-0.13}^{+0.17}$&$1.32_{-0.16}^{+0.21}$&$228_{-15}^{+13}$&$^{*}$&-&\\\noalign{\smallskip}
                        \hline
                \end{tabular}
        \end{center}
        $^{*}$: The spectral parameters are obtained from \citet{Kole2020}, who performed a joint fit using an external spectrum with POLAR data based on the Multi-Mission Maximum Likelihood (3ML) framework \citep{Vianello2015}.
\end{table*}

\begin{table*}[htbp]
        \caption{Spectral parameters and polarization properties of the 20 GRBs observed with AstroSat}
        \label{tab3: AstroSat}
        \begin{center}
                \begin{tabular}{c c c c c c c c c c}
                        \hline\hline\noalign{\smallskip}
                        GRB  &  $PD_{obs}$(\%)$$&$\alpha_{s}$&$\beta_{s}$&$E_{p,obs}$(keV)&Instrument(Spectrum)&$z$&\\
                        \hline\noalign{\smallskip}
                        160325A&$<45.02$&$0.25_{-0.08}^{+0.07}$&$0.97_{-0.10}^{+0.14}$&$223.57_{-25}^{+29}$&Fermi-GBM, BAT&$-$&\\[5pt]
                        160623A&$<56.51$&$-0.06_{-0.02}^{+0.02}$&$1.83_{-0.09}^{+0.10}$&$662_{-18}^{+19}$&Fermi-GBM, Konus-Wind&0.367&\\[5pt]
                        160703A&$<62.64$&$-0.22_{-0.12}^{+0.09}$&$1.48^{a}$&$351_{-46}^{+40}$&BAT, Konus-Wind&$-$&\\[5pt]
                        160802A&$<51.89$&$-0.36_{-0.04}^{+0.03}$&$1.53_{-0.14}^{+0.20}$&$207_{-1}^{+1}$&Fermi-GBM&$-$&\\[5pt]
                        160821A&$<33.87$&$-0.04_{-0.00}^{+0.00}$&$1.29_{-0.02}^{+0.02}$&$977_{-12}^{+12}$&Fermi-GBM, BAT&$-$&\\[5pt]
                        170527A&$<36.46$&$-0.01_{-0.01}^{+0.01}$&$2.14_{-0.29}^{+0.29}$&$974_{-47}^{+51}$&Fermi-GBM&$-$&\\[5pt]
                        171010A&$<30.02$&$0.12_{-0.01}^{+0.00}$&$1.39_{-0.02}^{+0.02}$&$180_{-3}^{+3}$&Fermi-GBM&$0.3285$&\\[5pt]
                        171227A&$<55.62$&$-0.20_{-0.01}^{+0.01}$&$1.49_{-0.05}^{+0.05}$&$899_{-32}^{+32}$&Fermi-GBM&$-$&\\[5pt]
                        180103A&$71.43_{-26.84}^{+26.84}$&$0.31_{-0.06}^{+0.06}$&$1.24_{-0.90}^{+0.13}$&$273_{-23}^{+26}$&BAT, Konus-Wind&$-$&\\[5pt]
                        180120A&$62.37_{-29.79}^{+29.79}$&$0.01_{-0.01}^{+0.01}$&$1.40_{-0.09}^{+0.09}$&$140.91_{-3}^{+3}$&Fermi-GBM&$-$&\\[5pt]
                        180427A&$60.01_{-22.32}^{+22.32}$&$-0.71_{-0.08}^{+0.08}$&$1.80_{-0.16}^{+0.16}$&$147_{-2}^{+2}$&Fermi-GBM&$-$&\\[5pt]
                        180806A&$<95.80$&$-0.08_{-0.04}^{+0.04}$&$1.46_{-0.23}^{+0.44}$&$453_{-44}^{+46}$&Fermi-GBM&$-$&\\[5pt]
                        180809B&$<24.63$&$-0.31_{-0.08}^{+0.07}$&$1.29_{-0.07}^{+0.08}$&$251_{-15}^{+16}$&BAT, Konus-Wind&$-$&\\[5pt]
                        180914A&$<33.55$&$-0.27_{-0.03}^{+0.03}$&$1.30_{-0.11}^{+0.15}$&$330_{-19}^{+20}$&Fermi-GBM&$-$&\\[5pt]
                        180914B&$48.48_{-19.69}^{+19.69}$&$-0.25_{-0.04}^{+0.04}$&$1.10_{-0.08}^{+0.70}$&$453_{-24}^{+26}$&BAT, Konus-Wind&$1.096$&\\[5pt]
                        190530A&$46.85_{-18.53}^{+18.53}$&$-0.01_{-0.02}^{+0.00}$&$2.50_{-0.25}^{+0.25}$&$888_{-8}^{+8}$&Fermi-GBM&0.9386&\\[5pt]
                        190928A&$<33.10$&$0.00_{-0.06}^{+0.06}$&$0.97_{-0.07}^{+0.13}$&$658_{-88}^{+111}$&Konus-Wind&$-$&\\[5pt]
                        200311A&$<45.41$&$-0.05_{-0.02}^{+0.02}$&$1.57_{-0.19}^{+0.19}$&$1218_{-110}^{+110}$&Fermi-GBM&$-$&\\[5pt]
                        200412A&$<53.84$&$-0.30_{-0.05}^{+0.05}$&$1.50_{-0.21}^{+0.21}$&$256_{-7}^{+8}$&Fermi-GBM&$-$&\\[5pt]
                        200806A&$<54.73$&$-0.47$&$1.96$&$109.12$&BAT&$-$&\\\noalign{\smallskip}
                        \hline
                \end{tabular}
        \end{center}
        $^{a}$: Fitting this spectrum with the Band function only presents a lower limit on $\beta_s$ of 1.48.
\end{table*}

\citet{Yonetoku2011, Yonetoku2012} reported polarization observations of the prompt emission of GRB 100826A, GRB 110301A, and GRB 110721A with GAP. For GRB 100826A, an averaged polarization of $27\pm11\%$ with a confidence level of $99.4\%$ (2.9$\sigma$) was reported, with systematic errors being considered for the first time in their analysis. For GRB 110301A and GRB 110721A, the observed linear polarizations are $70\pm22\%$ and $84_{-28}^{+16}\%$ with confidence levels of $3.7\sigma$ and $3.3\sigma$, respectively. \citet{Berger2011} reported a redshift value of $0.382$ for GRB 110721A. The spectral parameters used in our calculations for all three GRBs are from the Fermi-GBM catalog in the energy range of $50$ keV$-300$ keV\footnote{https://heasarc.gsfc.nasa.gov/W3Browse/fermi/fermigbrst.html} \citep{Kienlin2020ApJ, Gruber2014, Kienlin2014ApJ, Bhat2016}, and are presented in Table \ref{tab1: GAP}.

Recently, \citet{Kole2020} published a detailed polarization catalog reporting the polarization properties of 14 GRBs observed with POLAR. We searched the parameters of the Band function for all of these GRBs and list them in Table \ref{tab2: POLAR} along with the instruments that provide them. Among these spectral parameters, all those from Konus were measured in the energy range of 20 keV$-$15 MeV \citep{Frederiks2016, Tsvetkova2017, Svinkin2017}, while those from Fermi-GBM were obtained in the energy range of 50 keV$-$300 keV \citep{Kienlin2017GCN, Roberts2017GCN, Stanbro2017GCN}.

The updated polarization measurements and the corresponding spectral parameters of 20 GRBs observed by the CZTI on board AstroSat were also reported by \citet{Chattopadhyay2022} recently. In Table \ref{tab3: AstroSat} we list the detailed polarization information and spectral properties for them \citep{Chattopadhyay2022}. In addition, the redshift values have been reported for 4 of 20 GRBs \citep{Malesani2016, Postigo2017, Gupta2022, GCN23246}, which provides more precise parameters for our calculations.

\section{Model and numerical results}
An ultrarelativistic jet is assumed to be an optically thin shell to $\gamma$-rays with an emitting region of radius $r$, located at redshift $z$, and a source with a luminosity distance $d_L$. Its fluence can be expressed as follows \citep{Toma2009, Granot1999, Woods1999, Ioka2001}:
\begin{equation}
        F=\frac{1+z}{d_L^2}r^2\int_{\nu_1}^{\nu_2}d\nu_{obs}\int_0^{\theta_j+\theta_V}\frac{f(\nu')d(\cos\theta)}{\gamma^2(1-\beta_0\cos\theta)^2}
        \int_{-\Delta\phi}^{\Delta\phi} A_0d\phi.
\end{equation}
In the above equation, $\theta_V$ is the viewing angle of the observer, $\theta_j$ is the half-opening angle of the jet, and $\theta$ is the angle between the line of sight and the local direction of the fluid velocity. The physical quantities that are primed and unprimed are in the comoving and observer frame, respectively. For example, $\nu'=\nu_{obs}(1+z)\gamma(1-\beta_0\cos\theta)$ is the observational frequency in comoving frame, with the bulk Lorentz factor $\gamma$ and the velocity of jet $\beta_0$ in units of the speed of light. $\nu_{obs}$ is the observationl frequency in the observer frame. $\nu_1$ and $\nu_2$ are the energy ranges of the corresponding detectors (e.g., $\nu_1=50$ keV and $\nu_2=500$ keV for POLAR). $\phi$ is the angle between the projection of the jet axis and the projection of the local fluid velocity direction on the sky plane. More information about $\Delta\phi$ can be obtained from \citet{Toma2009}. $E_{p,obs}$ can be converted into the comoving frame by $E'_p=E_{p,obs}(1+z)/2/\gamma$. We adopted the following form for the spectrum of GRB prompt emission described by the Band function \citep{Band1993}:\begin{equation}
        f(\nu')=\begin{cases}
                \vspace{1ex}
                {(\frac{\nu'}{\nu'_0})}^{-\alpha_s}e^{-{\frac{\nu'}{\nu'_0}}}, & \text{$\nu'<\nu'_0(\beta_s-\alpha_s)$}, \\ {(\frac{\nu'}{\nu'_0})}^{-\beta_s}(\beta_s-\alpha_s)^{\beta_s-\alpha_s}e^{\alpha_s-\beta_s}, & \text{$\nu'\geq\nu'_0(\beta_s-\alpha_s)$}.
        \end{cases}
\end{equation}
$\alpha_s$ and $\beta_{s}$ are the low-energy and high-energy spectral indices, respectively. $\nu'_0=E'_p/h$ is the comoving break energy of the Band spectrum. $h$ is the Planck constant. In this paper, $\alpha_s$ and $\beta_s$ are the spectral indices of the flux desity $F_{\nu}$. In our calculation, the source was assumed to be at a redshift of 1 unless its redshift value has been reported. We assumed an aligned large-scale ordered magnetic field in the emission region with an orientation of $\delta=\pi/6$ \citep{Lan2016}. Other fixed parameters are $\theta_j=0.1$ rad, $\theta_V=0$ rad, and $\gamma=100$.

We then calculated the PDs of the GRBs with the polarization observation, using the observed spectral parameters (including $\alpha_s$, $\beta_s$ , and $E_{p,obs}$) and redshift values as well as the energy range of the polarimeters. In general, the calculated PD ($PD_{cal}$) of a GRB consists of a typical value and its upper and lower limits. In our calculations, we used the redshift value, the detector energy range, and the typical values of $E_{p,obs}$, $\alpha_s$, and $\beta_s$ to calculate a typical value of $PD_{cal}$. For the same GRB (i.e., the redshift value and the upper and lower limits of the detector energy range are fixed), the upper limit of its $PD_{cal}$ was taken when its $\alpha_s$ and $\beta_s$ took the maximum values and $E_{p,obs}$ took the minimum value; conversely, the minimum values of $\alpha_s$ and $\beta_s$ and the maximum value of $E_{p,obs}$ determine the lower limit of the $PD_{cal}$. We compare the calculated PDs ($PD_{cal}$) and the observed PDs ($PD_{obs}$) in Figs. 1-3.

\begin{figure}[htbp]
        \includegraphics[width=0.46\textwidth,height=0.36\textwidth]{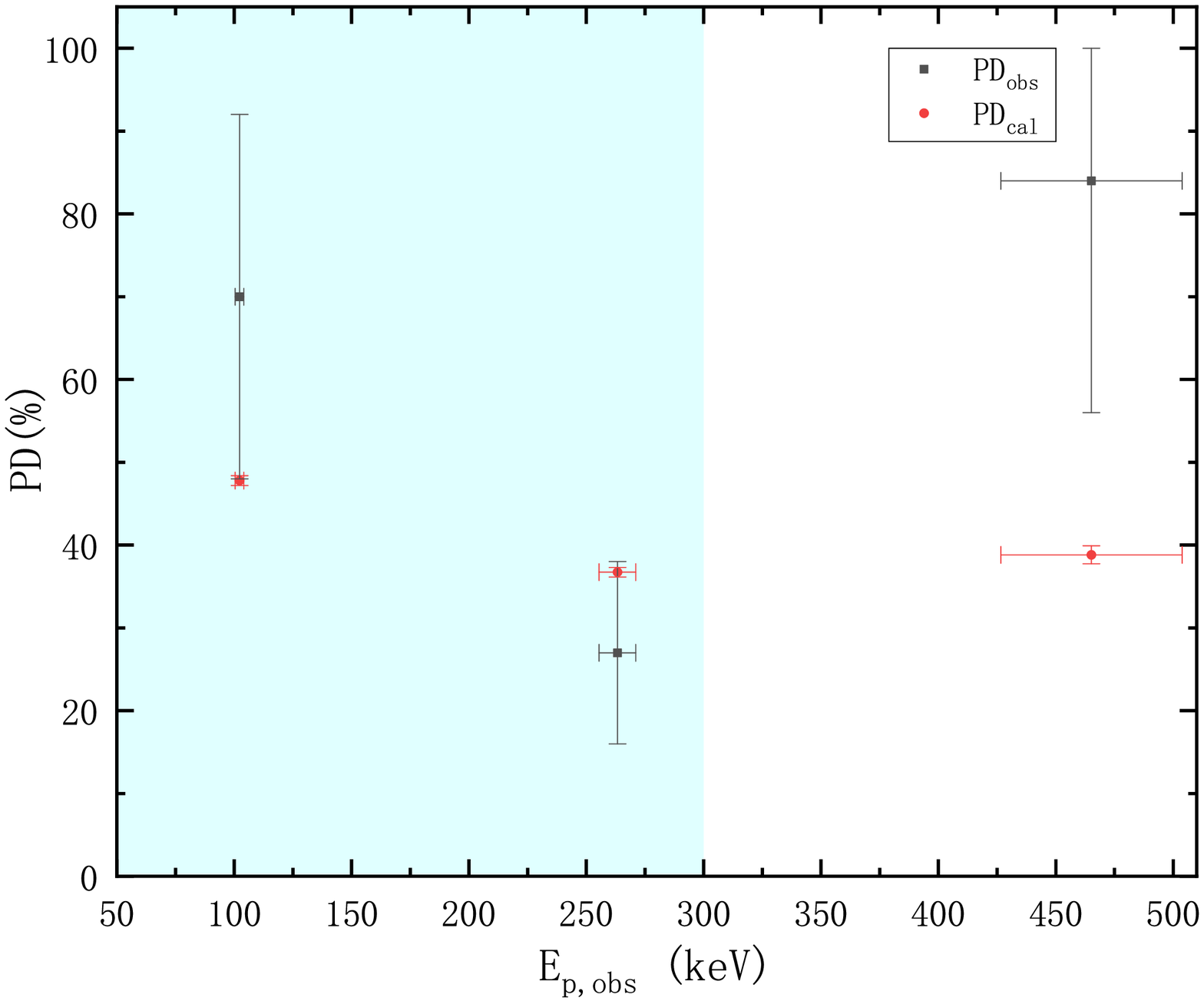}
        \caption{Ranges of $PD_{cal}$ and $PD_{obs}$ observed by GAP. The black diamond and red points correspond to the observed and calculated PD values, respectively. The area in light blue in the figure indicates the energy range of GAP, which is $50$ keV$-300$ keV.} \label{fig1: GAP}
\end{figure}

\begin{figure}[htbp]
        \includegraphics[width=0.47\textwidth,height=0.37\textwidth]{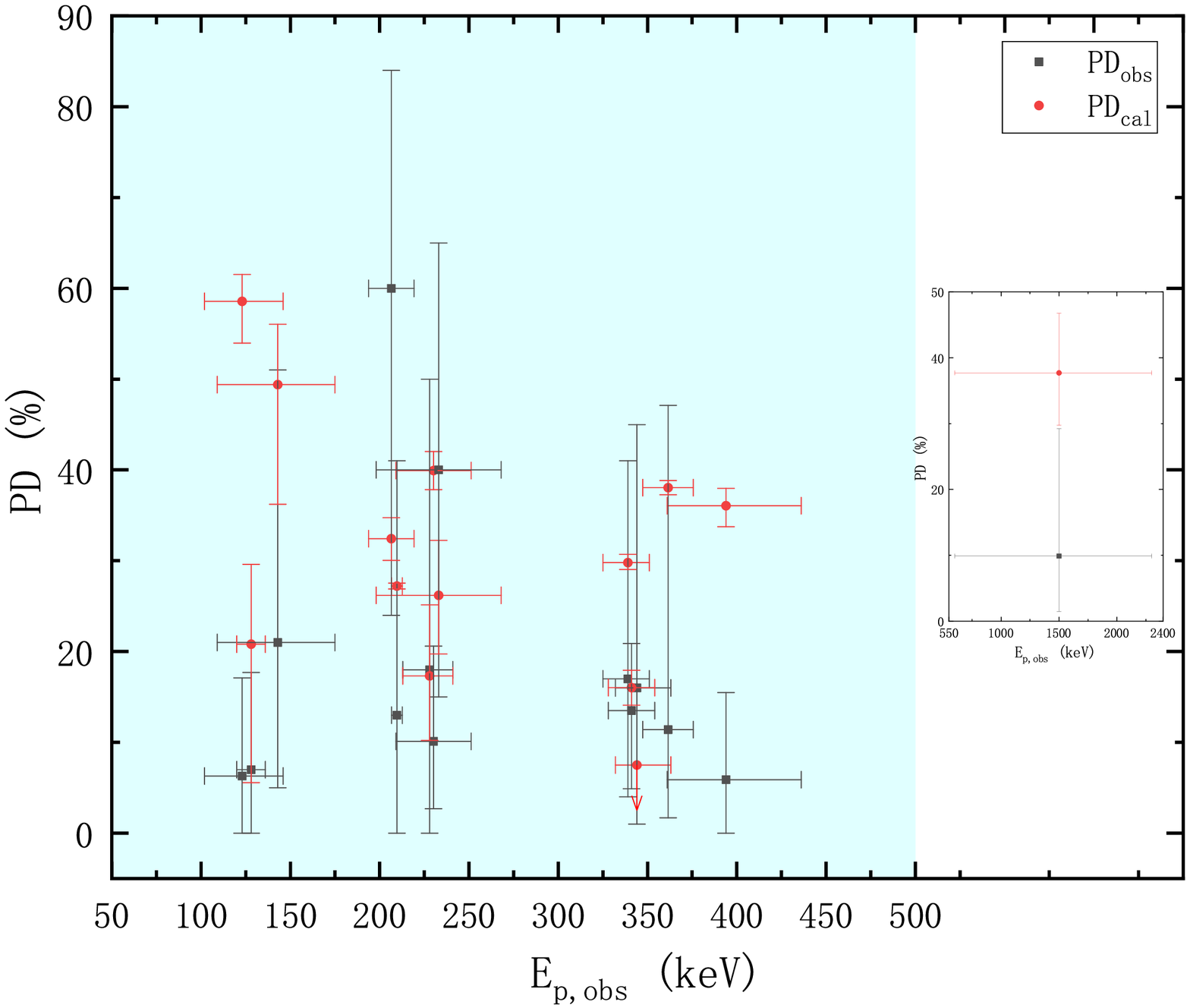}
        \caption{Same as Fig. 1, but for POLAR. The energy range of POLAR is $50$ keV$-500$ keV.} \label{fig2: POLAR}
\end{figure}

Fig. \ref{fig1: GAP} shows a comparison of $PD_{cal}$ and $PD_{obs}$ observed with GAP. The energy ranges of GAP (for polarization observations) and of the Fermi-GBM (for spectra observations) overlap exactly ($50$ keV$-300$ keV). For GRB 100826A detected with GAP, the polarization evolutions of this burst were simulated with the collision-induced magnetic reconnection model \citep{Deng2016}, and the results can reproduce the
time-resolved polarizations, especially the 90-degree polarization angle (PA) change between the two pulses. The observed PD of GRB 110721A is larger than the predicted one.
We also calculated the $PD_{cal}$ ranges and compared them with the $PD_{obs}$ observed by POLAR, as shown in Fig. \ref{fig2: POLAR}.  $PD_{cal}$  for 10 of the 13 GRBs in the light blue region overlaps with their corresponding $PD_{obs}$. The $PD_{cal}$ of the remaining 4 GRBs is significantly higher than the $PD_{obs}$ ranges. All these $PD_{cal}$ show higher polarization degrees.

\begin{figure}[htbp]
        \includegraphics[width=0.47\textwidth,height=0.37\textwidth]{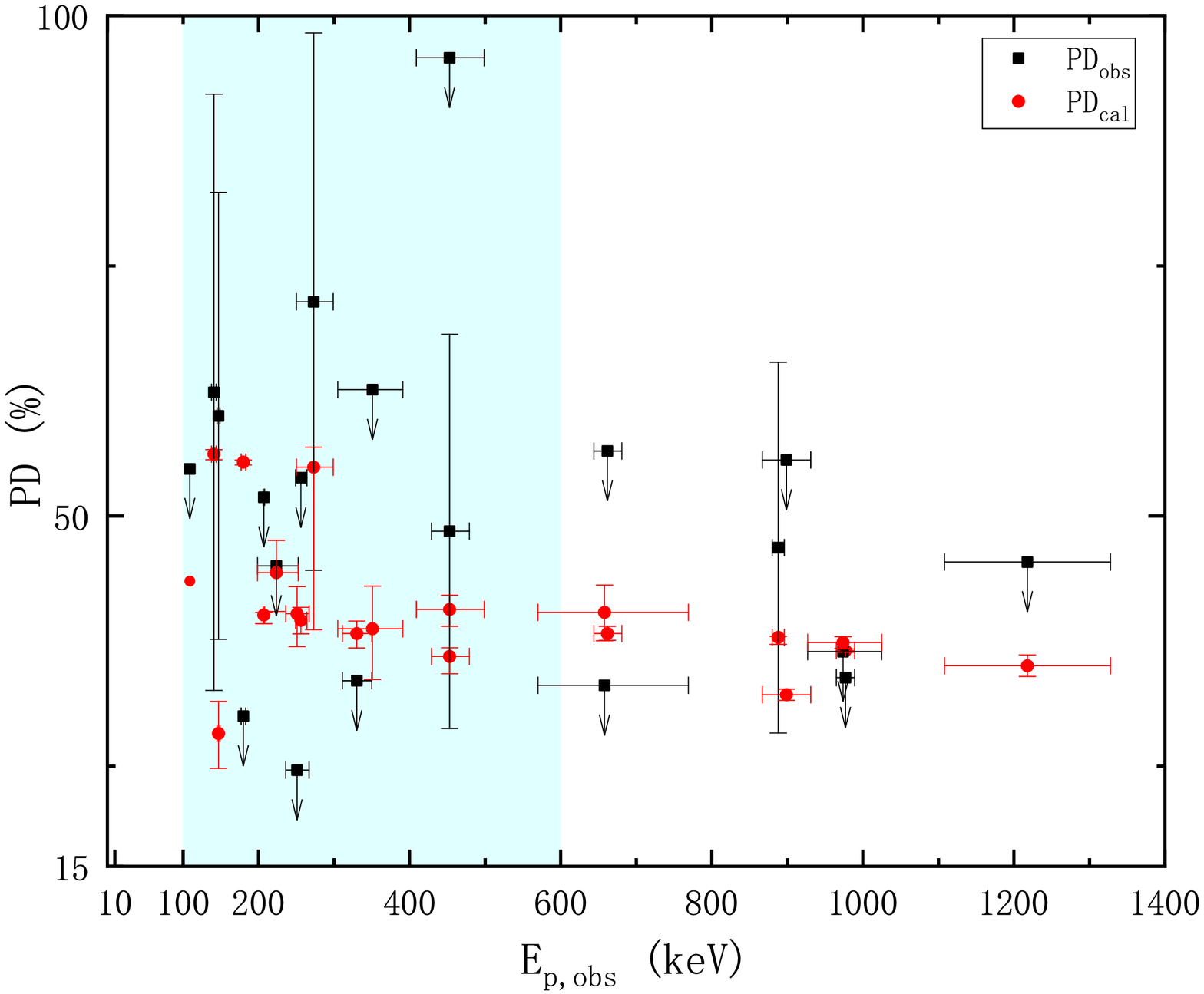}
        \caption{Same as Fig. 1, but for AstroSat. The energy range of AstroSat is $100$ keV$-600$ keV.} \label{fig3: AstroSat}
\end{figure}

In Fig. \ref{fig3: AstroSat} we numerically calculated the $PD_{cal}$ ranges of the GRBs observed by the AstroSat and found that the results match most of the observations, with a distribution around $\sim40\%$. The only burst with observed PD larger than the predicted value is GRB 180427A. Our integrated energy range of Stokes parameters for AstroSat is $100$ keV$-600$ keV. The range of $\Pi_0$ is $[0,1]$. This requires that the spectral index ($\alpha_s$ or $\beta_s$) should be higher than $-1$ according to the local polarization as shown below (Toma et al. 2009).

\begin{equation}
        \Pi_0\equiv
        \begin{cases}
                \vspace{1ex}
                {\displaystyle\frac{\alpha_s+1}{\alpha_s+\frac{5}{3}}},\text{$\nu'\leq\nu'_0(\beta_s-\alpha_s)$}\\
                {\displaystyle\frac{\beta_s+1}{\beta_s+\frac{5}{3}}},\text{$\nu'\geq\nu'_0(\beta_s-\alpha_s)$},
        \end{cases}
\end{equation}

The PDs of the POLAR bursts are concentrated around $10\%$, while they are around $40\%-50\%$ for AstroSat bursts. To interpret this discrepancy, we plot the spectral indices with peak energy in Fig. 4. The typical values of high-energy spectral indices are similar for POLAR bursts and AstroSat bursts. However, the typical value of low-energy spectral index is higher for AstroSat bursts (typically $\alpha_s\sim0.0$) than the POLAR bursts (typically $\alpha_s\sim-0.5$), resulting in a higher $PD_{cal}$ for AstroSat bursts. And the integrated energy range ($100$ keV$-600$ keV) of AstroSat bursts, compared with that ($50$ keV$-500$ keV) of POLAR bursts, is shifted to the higher energy range. For the bursts with similar spectral parameters the contributions from the high-energy photons (with larger local PD) will be larger for AstroSat burst, which will lead to a higher energy-integrated $PD_{cal}$. These might be the main reasons for this discrepancy.
\begin{figure}[htbp]
        \includegraphics[width=0.47\textwidth,height=0.37\textwidth]{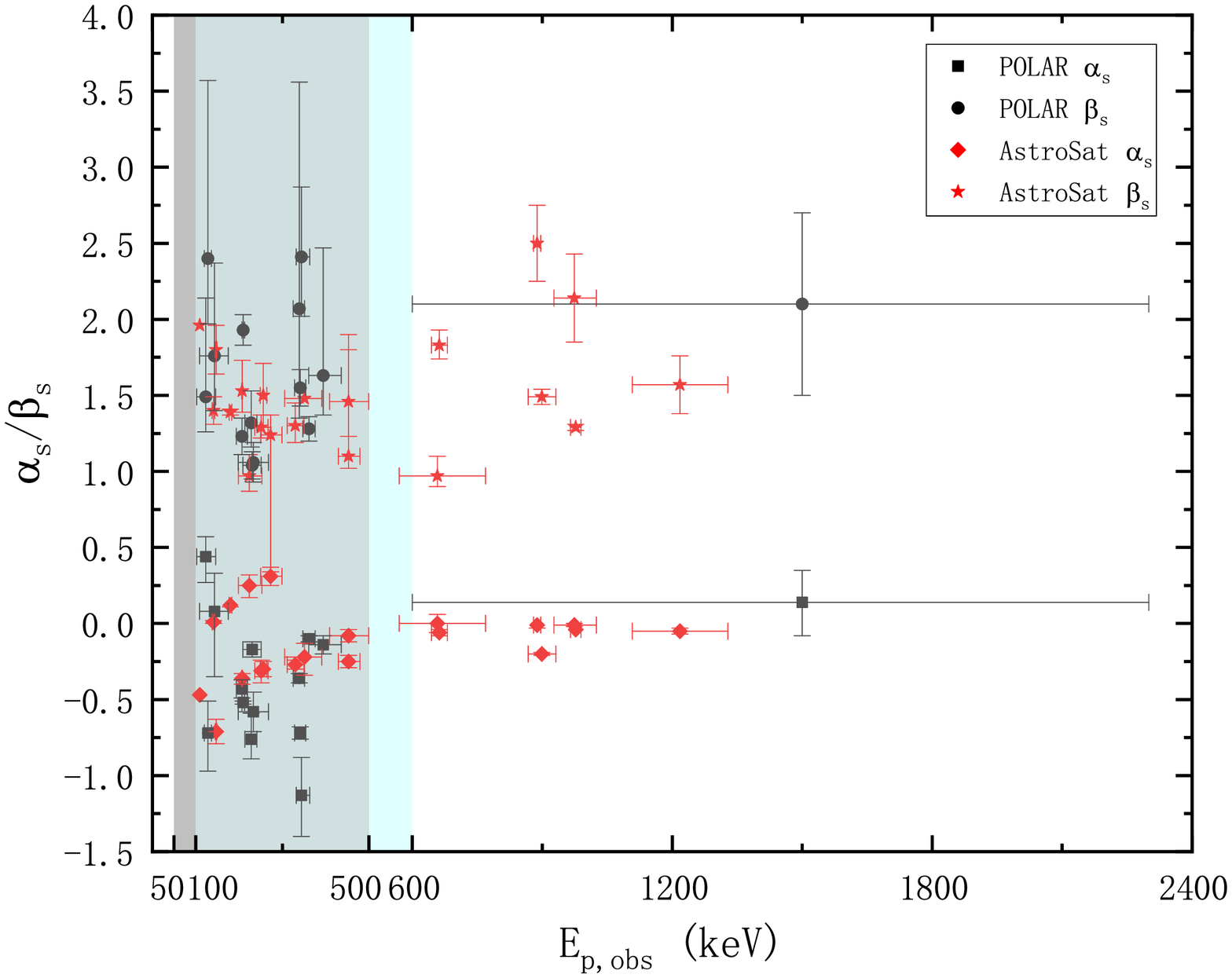}
        \caption{Comparison of the spectral parameters of the bursts observed with POLAR and AstroSat. The black squares and circles represent low- and high-energy spectral indices for bursts detected by POLAR, the red diamonds and stars correspond to low- and high-energy spectral indices for bursts observed by AstroSat.}\label{fig4: HPD and LPD}
\end{figure}

Because the energy range of the polarimeter also affects the observed polarization properties, we numerically predict the PDs of the long and short GRBs measured by two detectors, Low-energy Polarimetry Detector (LPD) and High-energy Polarimetry Detector (HPD) (whose energy ranges are $2$ keV$-30$ keV and $30$ keV$-800$ keV, respectively) on board POLAR-2 \citep{POLAR2}, based on the typical values and distribution of their spectral parameters \citep{Nava2011, Preece2000}. We present the results in Fig. \ref{fig4: HPD and LPD}, where the gray area and the light blue area denote the energy ranges of LPD and HPD, respectively. The typical PD values of GRBs detected by LPD and HPD are shown as black diamonds and red points. Because the typical PD values for different detectors are calculated under the typical spectral parameters and the number of GRBs with typical spectral parameters is maximum, the observed PD value with the maximum number of GRBs for each detector is concentrated around the typical PD values predicted by the model.
\begin{figure}[htbp]
        \includegraphics[width=0.47\textwidth,height=0.59\textwidth]{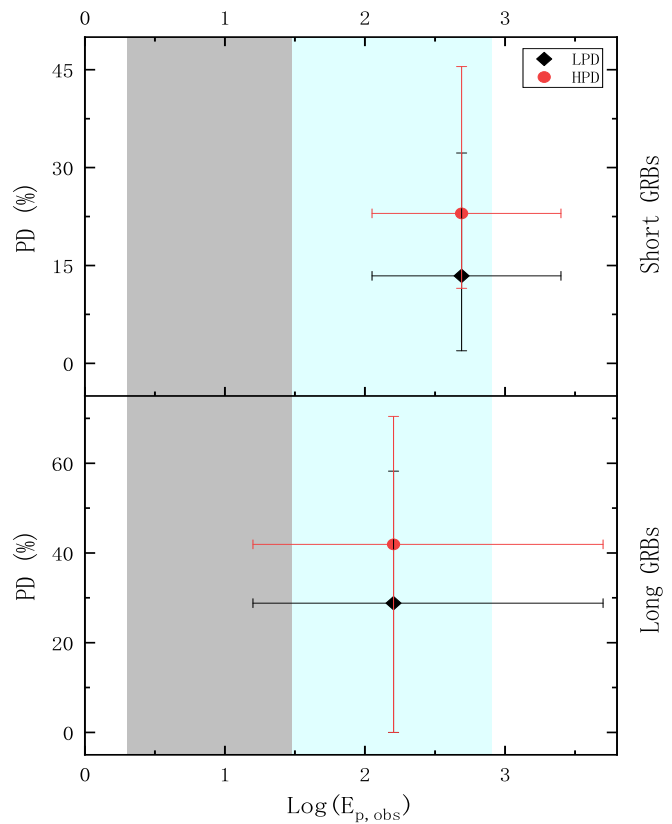}
        \caption{Comparison of polarization measurements of the long and short GRBs observed with HPD and LPD. The concentrated PD values of LPD and the HPD are shown with black diamonds and red points. }\label{fig4: HPD and LPD}
\end{figure}

\section{Discussion and conclusion}
Polarization properties of GRBs are essential for diagnosing the magnetic field configuration and the geometry of the emission region and the observation. We used the observed energy spectrum to calculate the corresponding GRB polarization properties within the synchrotron-emission model and compared them with the observed time-integrated PDs. In our model, we used the large-scale aligned magnetic field in the emission region. Therefore, the predicted PDs give upper limits for the synchrotron-emission models.

For GAP, POLAR, and AstroSat the predicted PDs of the model can match most of the corresponding observed PDs, indicating that in the GRB prompt phase or at least during the peak time of the burst, the magnetic field configuration is approximately large-scale ordered with $\xi_B>1$ \citep{Lan2019, Lan2021}. The large-scale ordered magnetic field in the GRB emission region may originate from its central engine. In the scenario of the internal shock in a fireball \citep{PX1994, RM1994}, the magnetic field may be mixed with a low $\xi_B$ value ($\xi_B<1$), so that this model is not favored by the current PD observations. For the internal shock with an ordered magnetic field \citep{Fan2004}, the magnetization parameter $\sigma$ is required to be smaller than 1. The observed PD values require that it cannot be too small, however, otherwise, turbulence will develop and destroy the ordered magnetic field \citep{Deng2017}. For the ICMART model \citep{Zhang2011}, the magnetic field becomes less ordered with the magnetic reconnection during the burst (i.e., a decrease in $\xi_B$). The observed data indicate that at the peak time of these bursts, the $\xi_B$ values of the magnetic fields are still higher than 1 (i.e., the magnetic fields are dominated by the ordered component at the peak times of the bursts).

For POLAR, $10\%$ of the observed PDs can also be interpreted as synchrotron emission in an ordered magnetic field with a small low-energy spectral index. However, $PD_{cal}$ of four GRBs is still higher than $PD_{obs}$. For these four GRBs, the magnetic field configurations in the emission regions may be mixed \citep{Lan2019}, or the PAs show rotation or an abrupt change during the bursts. Future time-resolved polarization observations will enable us to distinguish the two scenarios. For AstroSat, the predicted PDs are concentrated around $40\%$ and can interpret the measurements of all GRBs except one (GRB 180427A). There is a discrepancy between the moderately low PDs ($\sim10\%$) detected with POLAR and the relatively high PDs (about $40\%-50\%$) observed with AstroSat. The main reasons for this difference may originate from both the higher low-energy spectral indices and higher integrated energy range for AstroSat bursts.

The PD data of GRB 180427A detected by AstroSat and of GRB 110721A detected by GAP are both higher than the predicted values.  Therefore the two PD observations challenge the models invoking synchrotron radiation in an ordered magnetic field. Because the synchrotron radiation in an ordered magnetic field gives the upper limit of the PD of the synchrotron-emission models with a mixed magnetic field for on-axis observations \citep{Lan2020}, and the GRBs selected for the polarization analysis are usually bright (indicating on-axis observations), the PD data of GRB 180427A and GRB 110721A finally challenge the synchrotron-emission models for GRB prompt phase.

With co-observations of the HPD and LPD on board POLAR-2 \citep{POLAR2}, the polarization spectrum will be obtained in the near future. We predict that the concentrated PD values of the GRBs detected by the HPD will be higher than the LPD in the synchrotron-emission model. A reversed polarization spectrum was predicted by the dissipative photosphere model, however, that is, the concentrated PD values detected by the HPD will be lower than the LPD. The two models can be tested in this way with polarization observations of POLAR-2. The emission mechanism at the high-energy $\gamma-$ray band is multiple inverse-Compton scattering for the dissipative photosphere model \citep{Lundman2018}, which is different from the synchrotron-emission model. With the observations of the polarization spectrum of the POLAR-2, the emission mechanism in the high-energy $\gamma-$ray band can therefore be determined.

\begin{acknowledgements}
We thank the anonymous referee for his/her useful comments. We also thank Yan-Zhi Meng for useful discussions and Tanmoy Chattopadhyay for useful comment. This paper is dedicated to the 70th anniversary of the physics of Jilin University. This work is supported by the National Natural Science Foundation of China (grant No. 11903014).
\end{acknowledgements}
\vspace{-2mm}
\bibliography{ref}

\end{document}